\documentclass[twocolumn]{webofc}
\usepackage[varg]{txfonts}   % Web of Conferences font
\usepackage{units}   % Web of Conferences font
\newcommand{\dgr}{\ensuremath{^{\circ}}\xspace}

\begin{document}
\title{Covering the celestial sphere at ultra-high energies}
\subtitle{Full-sky cosmic-ray maps beyond the ankle and the flux suppression}

\author{\firstname{J.} \lastname{Biteau}\inst{1}\fnsep\thanks{\email{biteau(at)ipno.in2p3.fr}} \and
\firstname{T.} \lastname{Bister}\inst{2} \and
\firstname{L.} \lastname{Caccianiga}\inst{3} \and
\firstname{O.} \lastname{Deligny}\inst{1} \and
\firstname{A.} \lastname{di Matteo}\inst{4} \and
\firstname{T.} \lastname{Fujii}\inst{5} \and
\firstname{D.} \lastname{Harari}\inst{6} \and
\firstname{K.} \lastname{Kawata}\inst{5} \and
\firstname{D.} \lastname{Ivanov}\inst{7} \and
\firstname{J.P.} \lastname{Lundquist}\inst{7} \and
\firstname{R.} \lastname{Menezes de Almeida}\inst{8} \and
\firstname{D.} \lastname{Mockler}\inst{9} \and
\firstname{T.} \lastname{Nonaka}\inst{5} \and
\firstname{H.} \lastname{Sagawa}\inst{5} \and
\firstname{P.} \lastname{Tinyakov}\inst{4,10} \and
\firstname{I.} \lastname{Tkachev}\inst{10} \and
\firstname{S.} \lastname{Troitsky}\inst{10},
on behalf of the Pierre Auger and Telescope Array Collaborations}

\institute{Institut de Physique Nucl\'eaire d'Orsay (IPNO), Universit\'e Paris-Sud, Univ.\ Paris/Saclay, CNRS-IN2P3, Orsay, France \and
RWTH Aachen University, III. Physikalisches Institut A, Aachen, Germany \and
Universit\`a di Milano, Dipartimento di Fisica, Milano, Italy \and
Service de Physique Th\'eorique, Universit\'e Libre de Bruxelles, Brussels, Belgium \and
Institute for Cosmic Ray Research, University of Tokyo, Kashiwa, Chiba, Japan  \and
Centro At\'omico Bariloche and Instituto Balseiro (CNEA-UNCuyo-CONICET), San Carlos de Bariloche, Argentina  \and
High Energy Astrophysics Institute and Department of Physics and Astronomy, University of
Utah, Salt Lake City, Utah, USA  \and
Universidade Federal Fluminense, EEIMVR, Volta Redonda, RJ, Brazil  \and
Karlsruhe Institute of Technology, Institute for Experimental Particle Physics (ETP), Karlsruhe, Germany \and
Institute for Nuclear Research of the Russian Academy of Sciences, Moscow, Russia
}

\abstract{%
Despite deflections by Galactic and extragalactic magnetic fields, the distribution of ultra-high energy cosmic rays (UHECRs) over the celestial sphere remains a most promising observable for the identification of their sources. Thanks to a large number of detected events over the past years, a large-scale anisotropy at energies above 8\,EeV has been identified, and there are also indications from the Telescope Array and Pierre Auger Collaborations of deviations from isotropy at intermediate angular scales (about 20 degrees) at the highest energies. In this contribution, we map the flux of UHECRs over the full sky at energies beyond each of two major features in the UHECR spectrum -- the ankle and the flux suppression --, and we derive limits for anisotropy on different angular scales in the two energy regimes. In particular, full-sky coverage enables constraints on low-order multipole moments without assumptions about the strength of higher-order multipoles. Following previous efforts from the two Collaborations, we build full-sky maps accounting for the relative exposure of the arrays and differences in the energy normalizations. The procedure relies on cross-calibrating the UHECR fluxes reconstructed in the declination band around the celestial equator covered by both observatories. We present full-sky maps at energies above ${\sim}\,10\,$EeV and ${\sim}\,50\,$EeV, using the largest datasets shared across UHECR collaborations to date. We report on anisotropy searches exploiting full-sky coverage and discuss possible constraints on the distribution of UHECR sources.
}
\maketitle
\section{The quest for anisotropies at ultra-high energies}
\label{intro}

After over a decade of observation with the largest cosmic-ray observatories at ultra-high energies  (UHE, $E > 10^{18}{\rm\,eV}\equiv1{\rm\,EeV}$), the arrival directions of the most energetic particles known to date appear to be nearly isotropic, as measured from the Northern Hemisphere by the Telescope Array (TA) and from the Southern Hemisphere by the Pierre Auger Observatory (Auger). The large exposure cumulated by these two arrays over the years is nonetheless starting to enable a glimpse at preferred directions beyond the ankle of the cosmic-ray spectrum, which may just be the tip of the iceberg. 

At $E_{\rm Auger}>8{\rm\,EeV}$, the Pierre Auger Collaboration reported the detection of a large-angular scale modulation of the cosmic-ray event rate in right ascension, with a small but significant (${>}\,5\,\sigma$) amplitude of $4.7\pm0.8\,\%$  \cite{2017Sci...357.1266P,2018ApJ...868....4A}. At energies where the cosmic-ray (CR) flux is suppressed, $E_{\rm TA}>57{\rm\,EeV}$, the Telescope Array Collaboration reported an indication (${>}\,3\,\sigma$) of clustering of events on an intermediate angular scale of $20\,\dgr$ \cite{2014ApJ...790L..21A}. The most significant excess found with the Pierre Auger Observatory dataset, for threshold energies between $E_{\rm Auger}>40{\rm\,EeV}$ and $E_{\rm Auger}>80{\rm\,EeV}$, did not reach the $3\,\sigma$ level using the procedure adopted for model-independent searches \cite{2015ApJ...804...15A}.

\begin{figure*}
\centering
\includegraphics[width=0.49\textwidth]{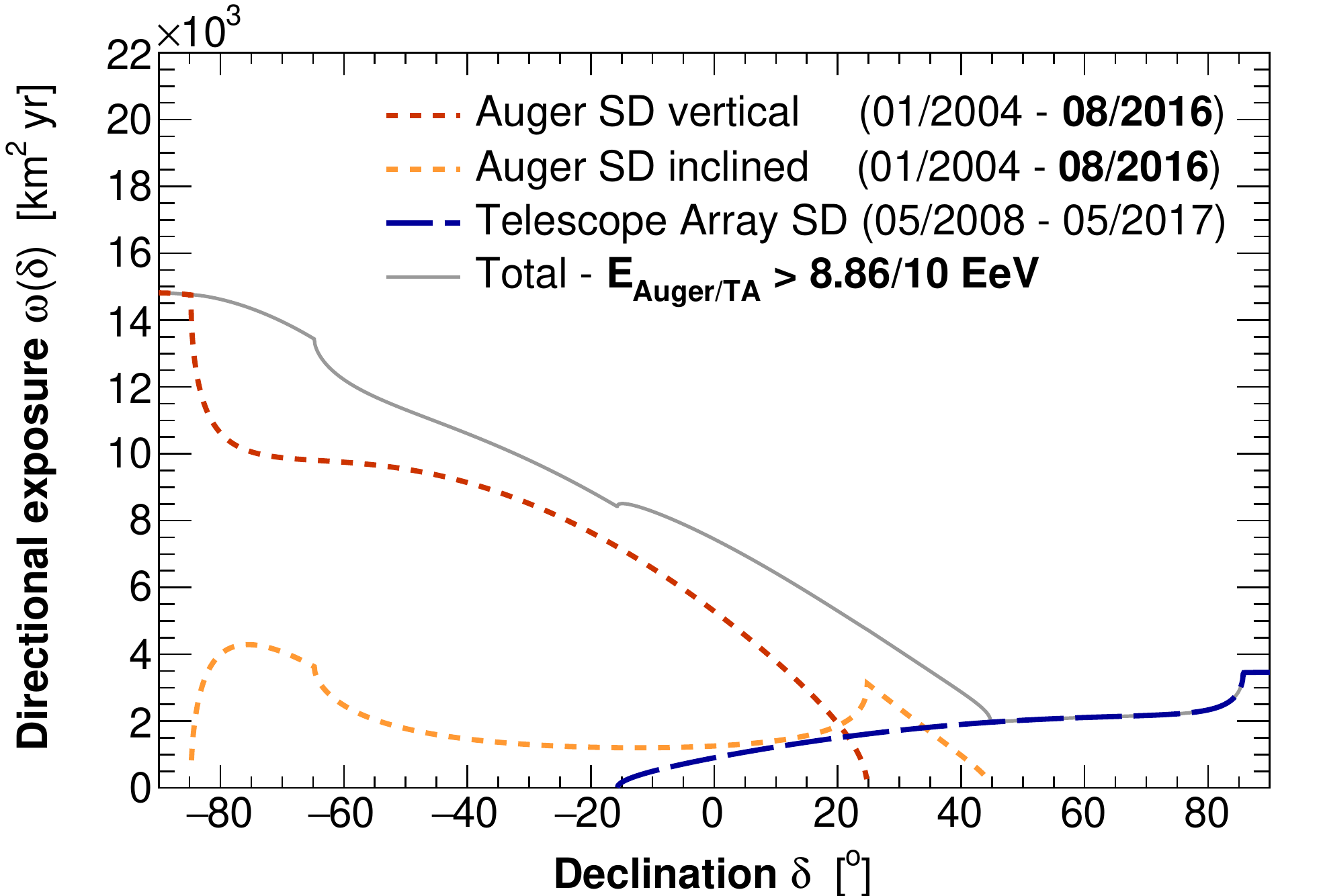}\hfill\includegraphics[width=0.49\textwidth]{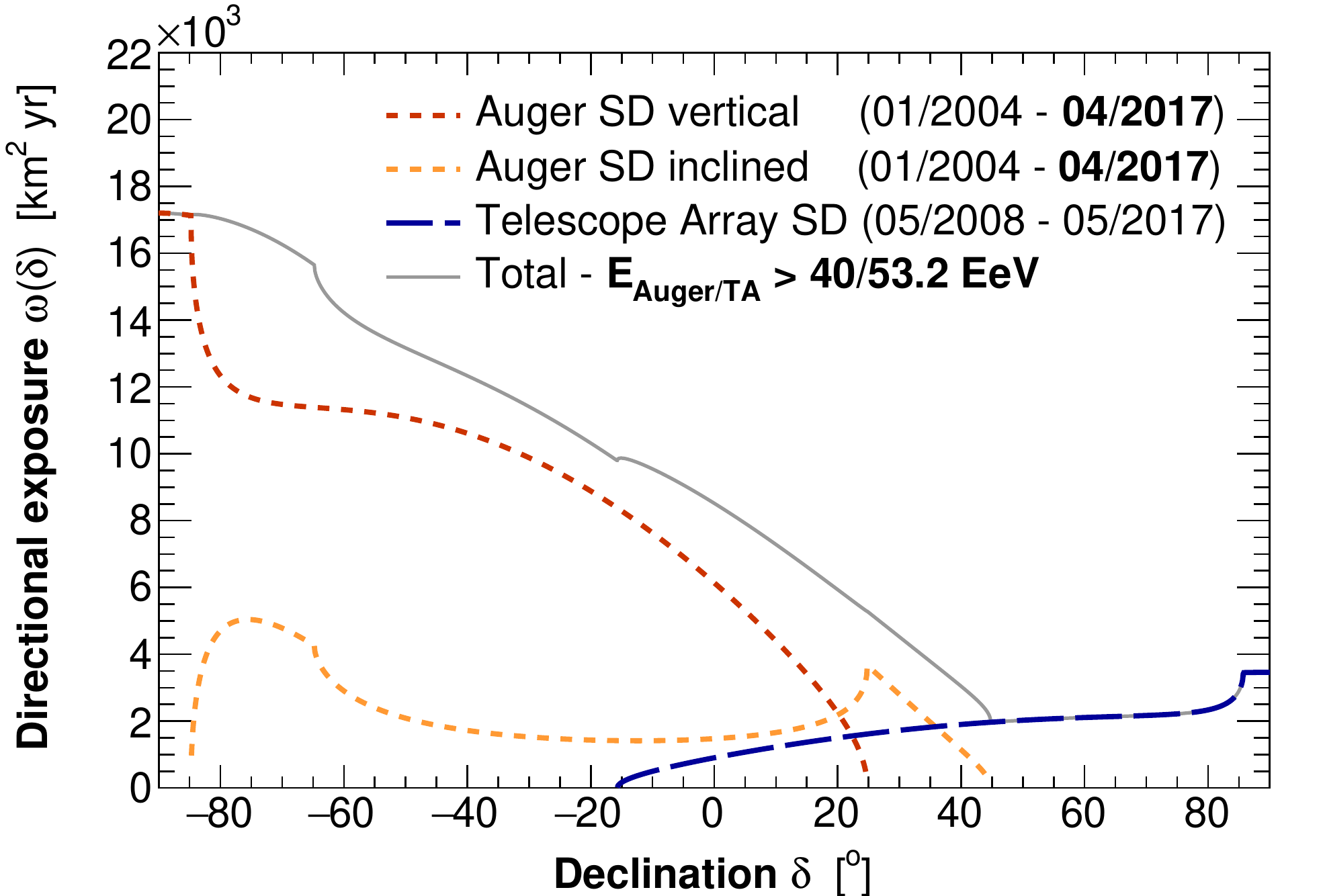}
\caption{Directional exposure from the surface detectors of the Telescope Array (long-dashed blue line: events  up to zenith angles of $55\,\dgr$) and Pierre Auger Observatory (dashed red line: vertical events up to zenith angles of $60\,\dgr$; dashed orange line: inclined events with zenith angles in $60-80\,\dgr$). {\it Left: } Dataset compiled for $E_{\rm Auger/TA} > \unit[8.86/10]{EeV}$. {\it Right: } Dataset compiled for $E_{\rm Auger/TA} > \unit[40/53.2]{EeV}$.}
\label{fig-exposure}       
\end{figure*}

The amplitude and direction of UHE anisotropies is expected to vary with energy. Observations beyond the ankle give access to a non-negligible fraction of the local Universe, {\it e.g.}\ up to 1\,Gpc for  cosmic rays with $E=10\,$EeV \cite{Aloisio:2017qoo}, while the horizon shrinks down to less than 100\,Mpc above 100\,EeV,  being limited by interactions with far-infrared and microwave diffuse photon fields  (the CIB and CMB). The highest-energy events possibly arise from a smaller number of sources and should suffer less from deflections within the Galactic magnetic field.\footnote{Median deflections are estimated to be on the order of ${\sim}\,3^\circ \times Z \times (E/100\,{\rm EeV})^{-1}$, with a spread of similar amplitude \cite{2016JCAP...01..037I}.} The study of anisotropies over the full celestial sphere in these regimes could thus provide invaluable insights into both the propagation of UHECRs over cosmic scales and the distribution of their sources, be they localized in a few directions or distributed following local structures.

\section{Ultra-high energy datasets}
\label{datasets}

Following previous joint searches over the full celestial sphere \cite{2014ApJ...794..172A,2018uhec.confa1020D}, we aim in this contribution at establishing and characterizing datasets covering energy ranges above the ankle and above the flux suppression of the cosmic-ray spectrum. In Sect.~\ref{datasetsTA-Auger}, we describe the datasets collected above threshold energies measured by the Telescope Array and the Pierre Auger Observatory, $E_{\rm TA}$ and $E_{\rm Auger}$, respectively. We combine in Sect.~\ref{flux-match} these datasets above common threshold energies, as defined from a match in flux in the declination band covered by both observatories.

\subsection{Telescope Array and Pierre Auger Observatory datasets}
\label{datasetsTA-Auger}

The Telescope Array has been fully operational since May 2008. Data with fiducial cuts described in  \cite{2018ApJ...867L..27A} were shared up to May 2017 above an energy threshold of $E_{\rm TA} = 10\,$EeV, where the array of 507 scintillator detectors is fully efficient.

The detectors of the Telescope Array pave an area of nearly 700\,km$^2$, with a zenith-angle coverage up to $55\,\dgr$. The geometrical exposure associated with the dataset studied in this contribution, corrected for bin-to-bin migration induced by the limited energy resolution, is estimated to $\unit[11{,}500]{km^{2}\,sr\,yr}$ at $E_{\rm TA} > 10\,$EeV, increasing by less than a percent at the highest energies ($\unit[11{,}600]{km^{2}\,sr\,yr}$ for $E_{\rm TA} > 50\,$EeV).

The Pierre Auger Observatory started taking data in 2004 and has been fully operational since January 2008. Two datasets were shared, with cuts optimized for, on the one hand, analyses above a full-efficiency threshold of $E_{\rm Auger} = 4\,$EeV \cite{2017Sci...357.1266P} and, on the other hand, analyses at the highest energies \cite{2018ApJ...853L..29A}.

Following \cite{2017Sci...357.1266P}, the low-energy dataset consists of events detected from 2004 January 1 up to 2016 August 31. Zenith angles up to $60\,\dgr$ are covered using so-called ``vertical'' events while a different reconstruction is used from $60\,\dgr$ up to $80\,\dgr$ for so-called ``inclined'' events. The $1{,}600$ water-Cherenkov detectors of the Pierre Auger Observatory are deployed over area of nearly 3{,}000\,km$^2$ resulting in a geometrical exposure amounting to $\unit[76{,}800]{km^{2}\,sr\,yr}$ for the period covered by the low-energy dataset. Correcting for energy resolution effects yields an unfolded exposure of $\unit[78{,}400]{km^{2}\,sr\,yr}$. At higher energies, above $E_{\rm Auger} = 40\,$EeV \cite{2015ApJ...804...15A}, the collected dataset covers the period 2004 January 1 up to 2017 April 30 as in \cite{2018ApJ...853L..29A}, with a geometrical exposure of  $\unit[86{,}900]{km^{2}\,sr\,yr}$, that is $\unit[91{,}300]{km^{2}\,sr\,yr}$ after correction for resolution effects.\footnote{The dataset is 2\,\% larger than that analyzed in \cite{2018ApJ...853L..29A}, both in terms of exposure and number of events, thanks to an improved data treatment.}

\begin{figure*}
\centering
\includegraphics[width=0.49\textwidth]{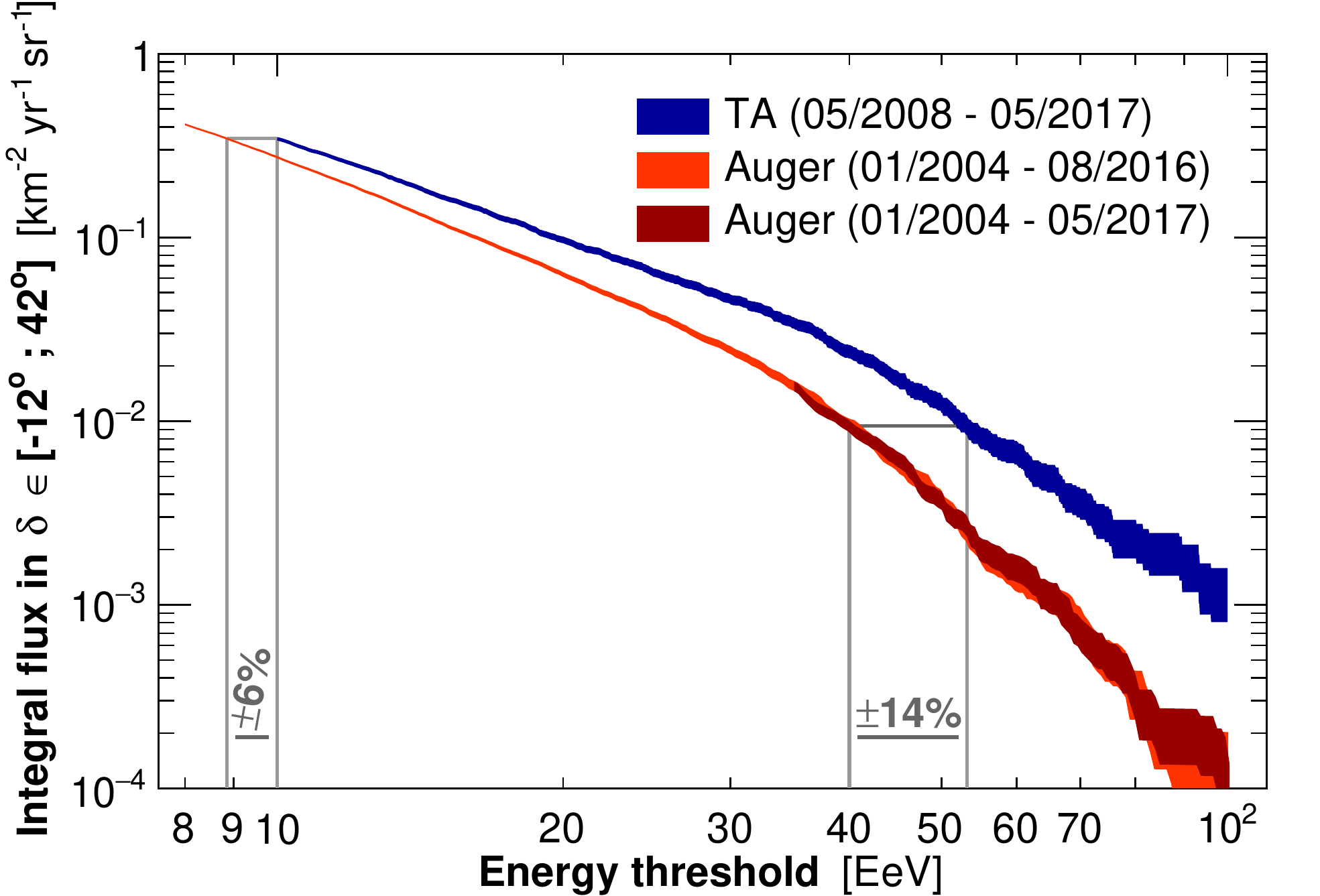}
\caption{Integral flux reconstructed in the common declination band as a function of energy threshold. The red and blue bands display the statistical uncertainty on the flux above threshold energies measured by the Pierre Auger Observatory and Telescope Array, respectively.}
\label{fig-integral-flux}       
\end{figure*}

\begin{table*}
\centering
\caption{Flux reconstructed in the declination band probed by both the Telescope Array and the Pierre Auger Observatory.}
\label{tab-flux}       
\begin{tabular}{lcc}
\hline
 $E_{\rm Auger/TA}$							& $\Phi_{\rm Auger}(>E_{\rm Auger/TA}, \delta\in[-12\,\dgr,24\,\dgr])$  & $\Phi_{\rm TA}(>E_{\rm Auger/TA}, \delta\in[-12\,\dgr,24\,\dgr])$\\
 											& [$\unit[]{km^{-2}\,yr^{-1}\,s^{-1}}$]  & [$\unit[]{km^{-2}\,yr^{-1}\,s^{-1}}$] \\
\hline
$\unit[8.86/10]{EeV}$						& $0.345\pm0.004$													& $0.345\pm0.008$ \\
$\unit[40/53.2]{EeV}$						& $(9.3\pm0.7)\times10^{-3}$										& $(9.6\pm1.2)\times10^{-3}$ \\
\hline
\end{tabular}
\end{table*}

\subsection{Flux match in the common declination band}
\label{flux-match}

The exposures from the Telescope Array and Pierre Auger Collaborations are determined with an accuracy better than 3\,\%, a level at which the contrast in flux reconstructed from the Northern and Southern hemispheres is not expected to have a significant impact on anisotropy constraints given current UHECR statistics. On the other hand, full-sky anisotropy studies could be substantially impacted by a systematic shift in the energy scales, the uncertainty on which is estimated to be 21\,\% and 14\,\% for the Telescope Array and Pierre Auger Observatory, respectively. Shifting the energy scale of a power-law spectrum of index $\Gamma$ by a factor $r$ results in a shift by a factor $r^{\Gamma-1}$ on the flux integrated above a fixed energy threshold. Thus, a $10-20\,\%$ mismatch in energy scale could result in a $30-60\,\%$ mismatch in flux for an index $\Gamma\sim4$. To account for such a flux mismatch, an analysis approach has been devised by the joint Working Group to ensure a proper cross-calibration of the datasets. The approach consists in comparing the flux reconstructed by each instrument in the fraction of the sky covered by both of them. The time-integrated exposure corresponding to each dataset being practically constant as a function of right ascension over the time period considered, a common declination band can be adopted to perform such a flux estimation. The overlap between the skies covered by the two observatories is illustrated in Fig.~\ref{fig-exposure}. Following \cite{2018uhec.confa1020D}, a common declination band ranging from $-12\,\dgr$ to $42\,\dgr$ is used to perform the cross-calibration of the datasets from the Pierre Auger Observatory and Telescope Array.

Because of different variations of the exposure as a function of declination for each observatory, a flux excess localized in the common band could cause different flux estimates. This would particularly be the case for a flux estimator built as $\int d\vec{n}\,N(\vec{n}) / \int d\vec{n} \,\omega(\vec{n})$, where $N$ and $\omega$ are the number of events and exposure, respectively, in the direction $\vec{n}$. Instead, an unbiased estimator of the flux can be quite naturally obtained as $\int d\vec{n}\,N(\vec{n}) / \omega(\vec{n}) = \sum_i 1/\omega(\vec{n_i})$, where $i$ indexes the event \cite{2018uhec.confa1020D}.

The unbiased estimator of the flux reconstructed by each observatory in the common declination band is shown in Fig.~\ref{fig-integral-flux} as a function of energy threshold. 
At low energies, driven by the full efficiency of the Telescope Array, we adopt a threshold $E_{\rm TA}>10\,$EeV. The flux reconstructed from the Telescope-Array dataset above this energy threshold is matched with the Auger dataset for an energy threshold $E_{\rm Auger}>8.86\,$EeV, as detailed in Table~\ref{tab-flux}. At higher energies, driven by the threshold adopted by the Pierre Auger Collaboration in model-independent anisotropy searches \cite{2015ApJ...804...15A} and by indications from model-dependent studies \cite{2018ApJ...853L..29A}, we reconstruct the flux in the common declination band above $E_{\rm Auger}>40\,$EeV. A compatible flux, reported in Table~\ref{tab-flux}, is obtained from the Telescope-Array dataset above an energy threshold $E_{\rm TA}>53.2\,$EeV. The amplitude of the mismatch in energy scale shown in Fig.~\ref{fig-integral-flux} is compatible with that reported in previous efforts, such as \cite{2014ApJ...794..172A} where a three times smaller dataset was investigated above $E_{\rm Auger/TA} > \unit[8.5/10]{EeV}$, and \cite{2018uhec.confa1020D} with a ${\sim}\,40\,\%$ smaller dataset above $E_{\rm Auger/TA}>\unit[42/57]{EeV}$.

The shift between the energy scales of the two observatories ranges from ${\pm}\,6\,\%$ up to ${\pm}\,14\,\%$, over the range of interest for this study. The absolute values are within systematic uncertainties reported by each Collaboration. The evolution of the shift with energy, that is the stretch in energy-scale, is estimated to ${\sim}\,10\,\%$ per energy decade. Energy-dependent systematic uncertainties have been investigated in the joint Working Group dedicated to the study of the cosmic-ray spectrum, concluding on estimations at the level of ${\sim}\,3\,\%$ and ${<}\,9\,\%$ per energy decade for the Auger and Telescope Array datasets, respectively.

\section{Search for anisotropies at all scales: analysis techniques}

Full-sky coverage provides access to an unbiased estimation of the angular power spectrum at all angular scales \cite{2014ApJ...794..172A}, particularly at low multipoles (dipole, quadrupole) where the full extent of anisotropy patterns can be probed without being impacted by the windowing effect resulting from partial coverage. Beyond a search based on a spherical-harmonics expansion, we also study, for the first time with full-sky coverage, localized excesses at the highest energies. Such searches are made possible by the reconstruction  of the UHECR flux and significance maps.

\subsection{Reconstruction of the flux and angular power spectrum}

As discussed in Sect.~\ref{flux-match} and \cite{2018uhec.confa1020D}, an unbiased estimator of the flux can be calculated as $\sum_i 1/\omega(\vec{n_i})$. Full-sky maps are built estimating the flux in circular windows centered on a HEALPix\footnote{\url{https://healpix.sourceforge.io/}} grid with $\texttt{nSide} = 64$. This parameter corresponds to an average pixel size of less than $1\,\dgr$, which matches the angular resolution of the instruments. Following previous searches \cite{2017Sci...357.1266P,2018uhec.confa1020D}, radii of $45\,\dgr$ and $20\,\dgr$ are adopted as default values for the datasets gathered at $E_{\rm Auger/TA} > \unit[8.86/10]{EeV}$ and $E_{\rm Auger/TA} > \unit[40/53.2]{EeV}$, respectively.

As for the flux, the angular power spectrum can be estimated from the inverse exposure as:
\begin{equation}
\begin{split}
C_\ell &= \frac{4\pi}{2\ell+1} \times \sum_m \left(\frac{a_{\ell m}}{a_{00}}\right)^2{,}\\
{\rm with\quad } a_{\ell m} &= \sum_i \frac{Y_{\ell m}(\vec{n_i})}{\omega(\vec{n_i})}{,}\nonumber
\end{split}
\end{equation}
where $Y_{\ell m}$ are the spherical harmonics.

Monte-Carlo simulations are used to estimate the distribution of the angular power that is expected to emerge from limited statistics, under the assumption of an isotropic distribution. The spread in the distribution of the angular power can be attributed to two sources:
\begin{itemize}
\item the limited number of events over the full sphere. This source of uncertainty can be singled out simulating a number of events identical to that gathered over each energy threshold according to the directional exposure presented in Fig.~\ref{fig-exposure};
\item the uncertainty in the energy cross-calibration procedure, due to the limited number of events in the cross-calibration band. The cumulated effect of this source of uncertainty and of the limited statistics over the full sphere can be estimated by varying, during the Monte Carlo simulation process, the normalization of the exposure of each observatory within an effective uncertainty. The main source of uncertainty is estimated from the resolution on the flux in the common band, as presented in Table~\ref{tab-flux}.\footnote{Multiple estimators were compared for the latter source of uncertainty and they were found to be consistent within rounding accuracy.}
\end{itemize}

%\newpage

\subsection{Quantifying the deviation from isotropy in all directions}

In addition to the flux evaluation, we aim at providing an estimation, over the sphere, of the uncertainty on the flux, whose variance is not only proportional to the flux but also to the exposure, larger in the Southern part of the sky, smaller in the Northern one.

More precisely, the significance of deviation from isotropy is estimated in each circular window centered over the HEALPix grid. The Li \& Ma estimator \cite{1983ApJ...272..317L} (Eq. 17) is used, defining each circular window as an ON-region and the rest of the sky as the OFF-region from which the background rate can be derived. The background rate is normalized by a factor, $\alpha$, computed as the ratio of the integral exposure over the ON- and OFF-regions. By convention, significances corresponding to deficits ($N_{\rm ON} < \alpha N_{\rm OFF}$) are quoted as negative values.  We checked that the presence of localized anisotropies in the OFF region does not affect in a significant manner the  Li \& Ma estimator. A comparison with a binomial estimator of the significance, such as used in \cite{2015ApJ...804...15A}, yields results consistent with the approach followed here. Overall, these tests all converge on an estimation of local significances with an accuracy on the order of $0.2-0.3\,\sigma$.

The procedure described above only provides estimates of local significances, not accounting for the blind search for flux excesses. The penalization procedure accounting for the search over the HEALPix grid, and over different search radii wherever relevant, is discussed in Sect.~\ref{interpretation}.

\section{Results with full-sky coverage}
\label{results}

The datasets gathered from the Pierre Auger Observatory and the Telescope Array encompass a total of ${\sim}\,31{,}000$ events at $E_{\rm Auger/TA} > \unit[8.86/10]{EeV}$ and ${\sim}\,1{,}000$ events at $E_{\rm Auger/TA} > \unit[40/53.2]{EeV}$, corresponding to an increase by ${+}\,200\,\%$ and ${+}\,60\,\%$ with respect to the datasets studied in \cite{2014ApJ...794..172A} and \cite{2018uhec.confa1020D}, respectively.

\subsection{Mapping the sky beyond the ankle}

\begin{figure*}
\centering
\includegraphics[width=0.72\textwidth]{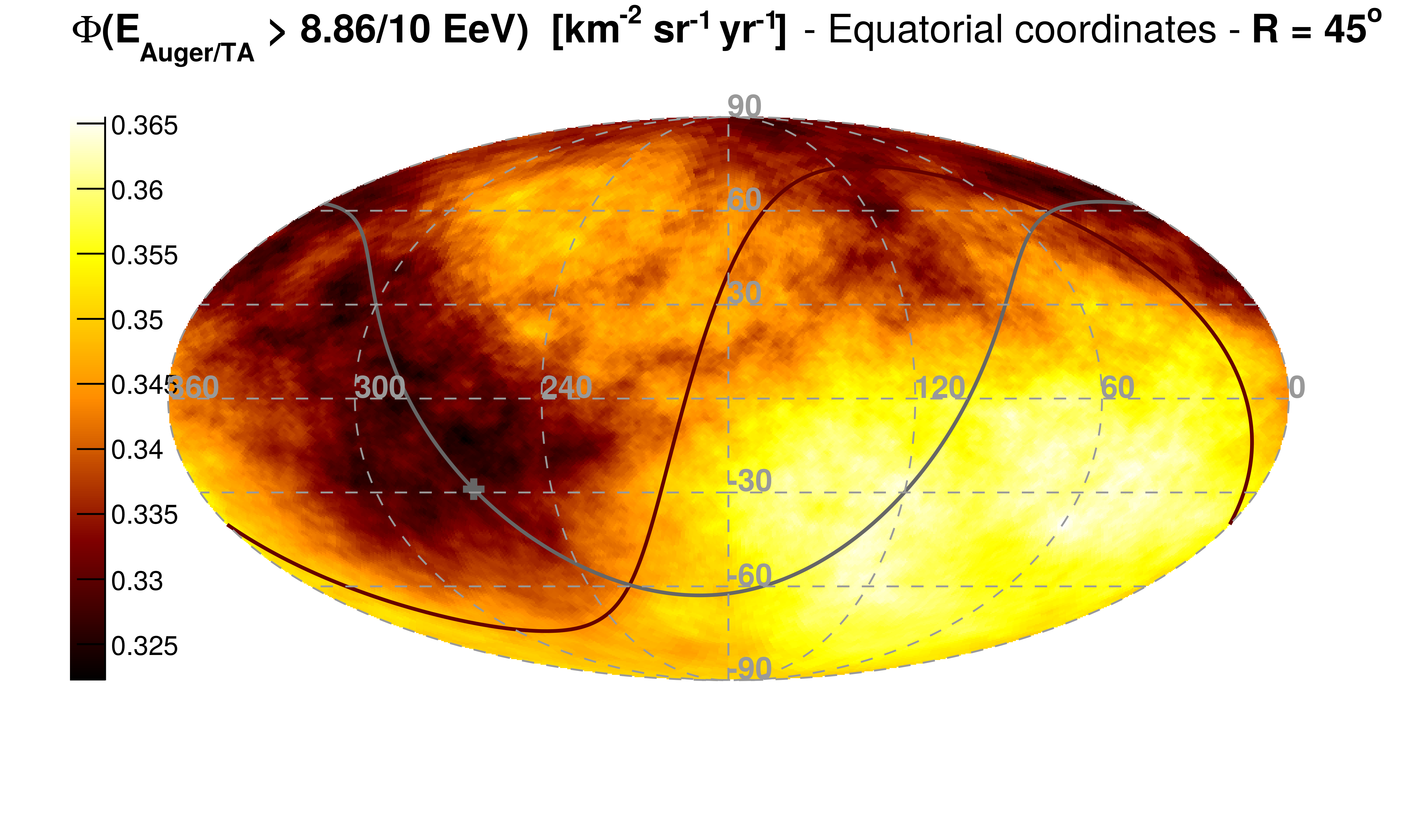}
\includegraphics[width=0.72\textwidth]{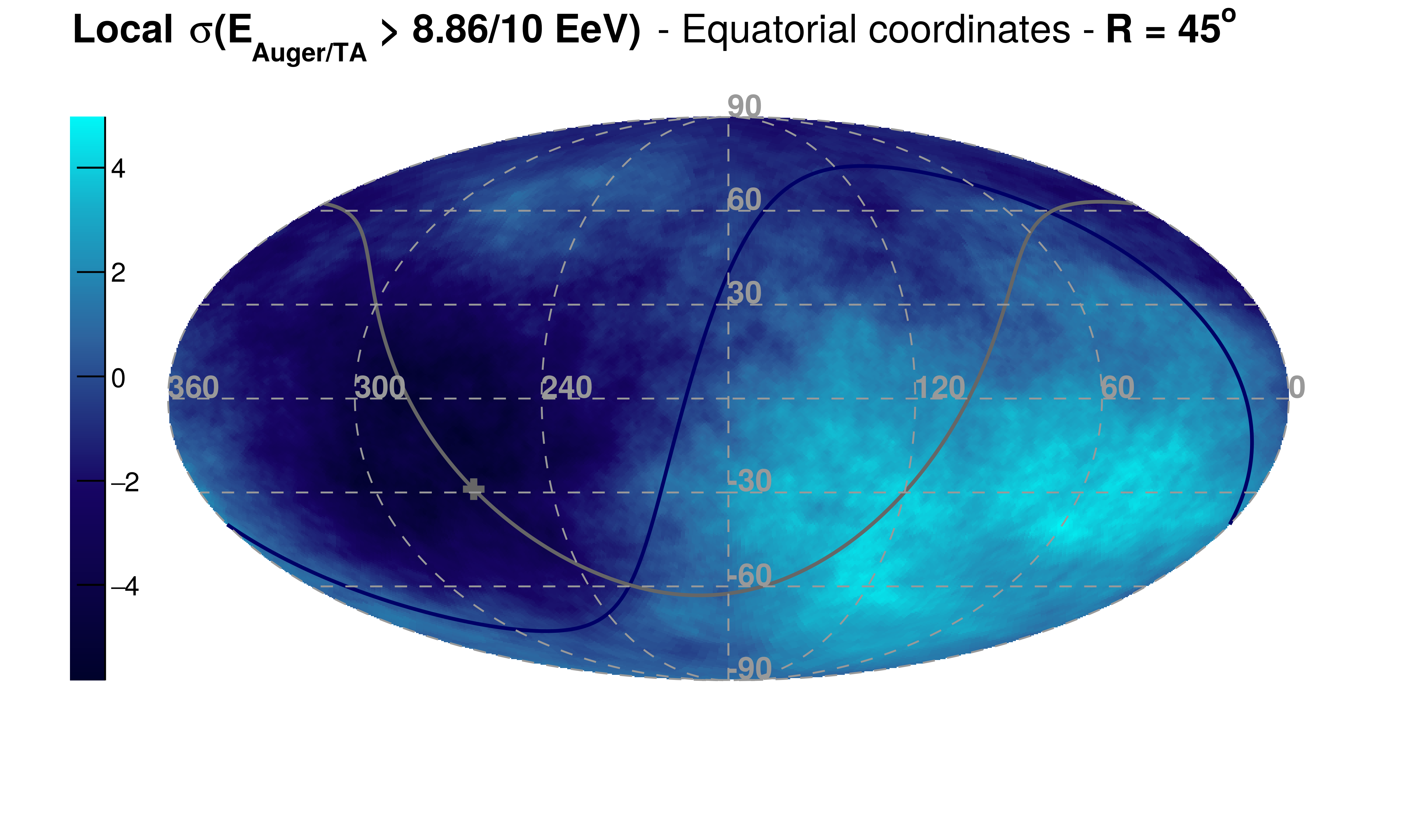}
\includegraphics[width=0.65\textwidth]{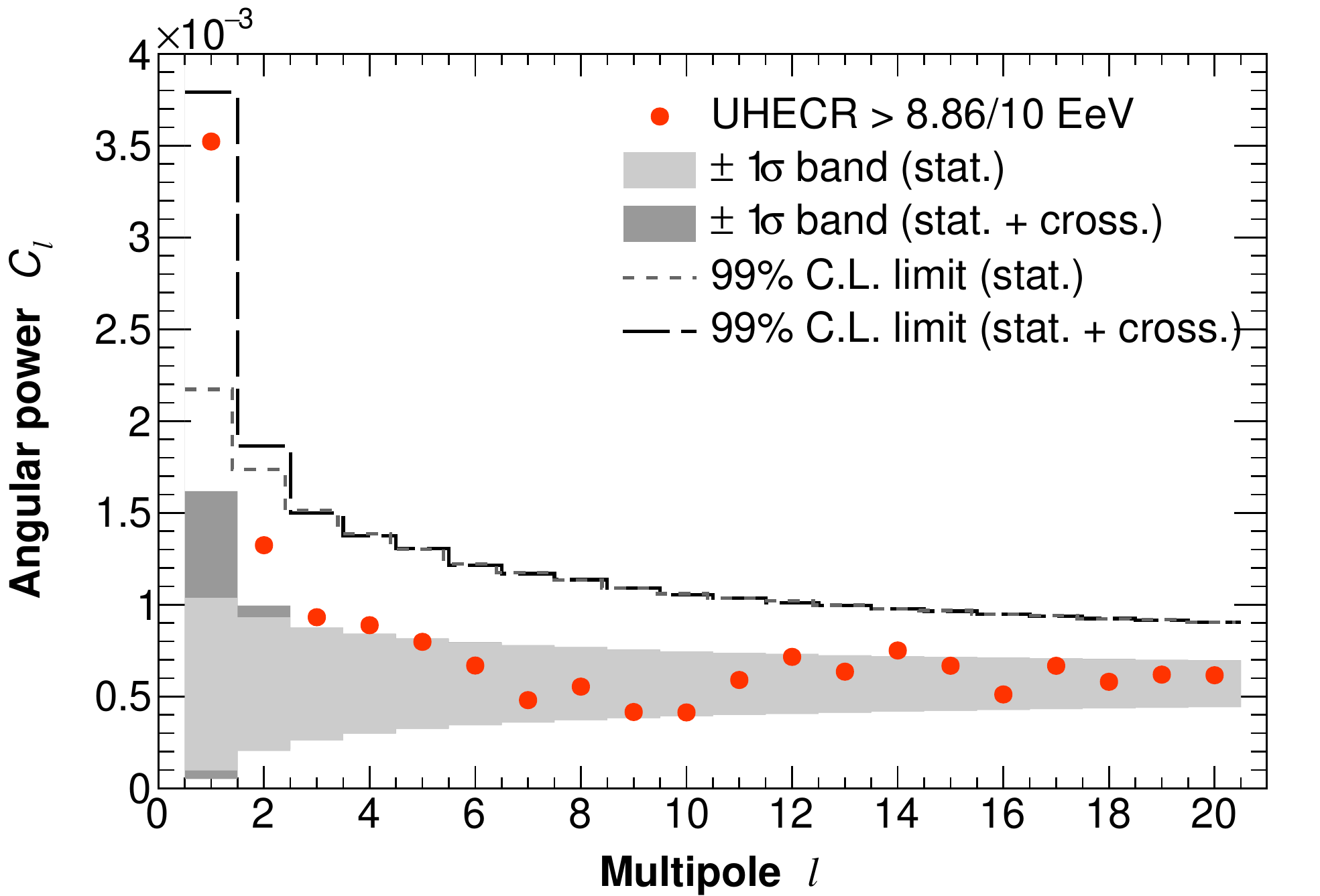}
\caption{Flux ({\it Top}) and significance ({\it Mid-panel}) maps reconstructed at $E_{\rm Auger/TA} > \unit[8.86/10]{EeV}$. The gray line indicates the Galactic plane and the dark line the supergalactic plane. {\it Bottom:} Angular power spectrum at $E_{\rm Auger/TA} > \unit[8.86/10]{EeV}$. The light-gray band indicates the ${\pm}\,\unit[1]{\sigma}$ confidence region for an isotropically distributed signal derived from the limited statistics over the full sphere. The dark-gray band further accounts for the uncertainty arising from the limited statistics in the cross-calibration band. Light-gray short-dashed and dark-gray long-dashed lines similarly indicate the 99\,\% confidence level upper limits derived under the assumption of an isotropic distribution.}
\label{fig-maps-lowE}       
\end{figure*}

The flux map reconstructed at $E_{\rm Auger/TA} > \unit[8.86/10]{EeV}$ is presented in equatorial coordinates in Fig.~\ref{fig-maps-lowE} (left), as estimated in $45\,\dgr$-radius windows. A large-scale feature is apparent from the enhanced flux in the hemisphere South of the supergalactic plane with respect to the North one. The presence of a smaller scale feature could be inferred from the median flux reconstructed in the North-West quadrant in equatorial coordinates. As presented in Fig.~\ref{fig-maps-lowE} (right), the significance of the North-West feature around $(\alpha,\delta) \sim (240\,\dgr, 60\,\dgr)$ is mild with respect to the minimum and maximum significances observed at declinations between ${\sim}\,-60\,\dgr$ and ${\sim}\,0\,\dgr$. This behavior can be understood from the decrease in exposure with increasing declination.

The angular power spectrum estimated at $E_{\rm Auger/TA} > \unit[8.86/10]{EeV}$ is shown in Fig.~\ref{fig-maps-lowE}, bottom. The largest deviation from isotropy is obtained for the dipole, at $\ell=1$, with a local significance of $\unit[2.5]{\sigma}$. As indicated by the dark- and light-gray bands, the uncertainty on the power at $\ell=1$ is largely driven by the limited number of events in the common declination band. Such an impact on the dipole can be expected from a varying contrast between the Northern and Southern hemispheres, which impacts the estimation of the component along the Earth rotation axis.

\begin{table*}
\centering
\caption{Constraints on the UHECR dipolar component along the Earth rotation axis, $d_z$, and along the perpendicular plane, $d_\perp$.}
\label{tab-dipole}       
\begin{tabular}{lcc}
\hline
 													& $d_z$ [\%] 					& $d_\perp$ [\%] 																		\\
\hline
Auger only, $\ell\leq1$ \cite{2017Sci...357.1266P}	& $-2.6 \pm 1.5$  & $6.0 \pm 1.0$ 																\\
Auger only, $\ell\leq2$  \cite{2018ApJ...868....4A} & $-2 \pm 4$ 			& $5.0 \pm 1.3$ 																 \\
Auger + TA (this work)				& $-2.6 \pm 1.3_{\rm stat} \pm 1.4_{\rm cross}$ & $4.3 \pm 1.1_{\rm stat} \pm 0.04_{\rm cross}$ \\\hline
\end{tabular}
\end{table*}

\subsection{Mapping the sky above the flux suppression}
\label{maps-highE}

\begin{figure*}
\centering
\includegraphics[width=0.72\textwidth]{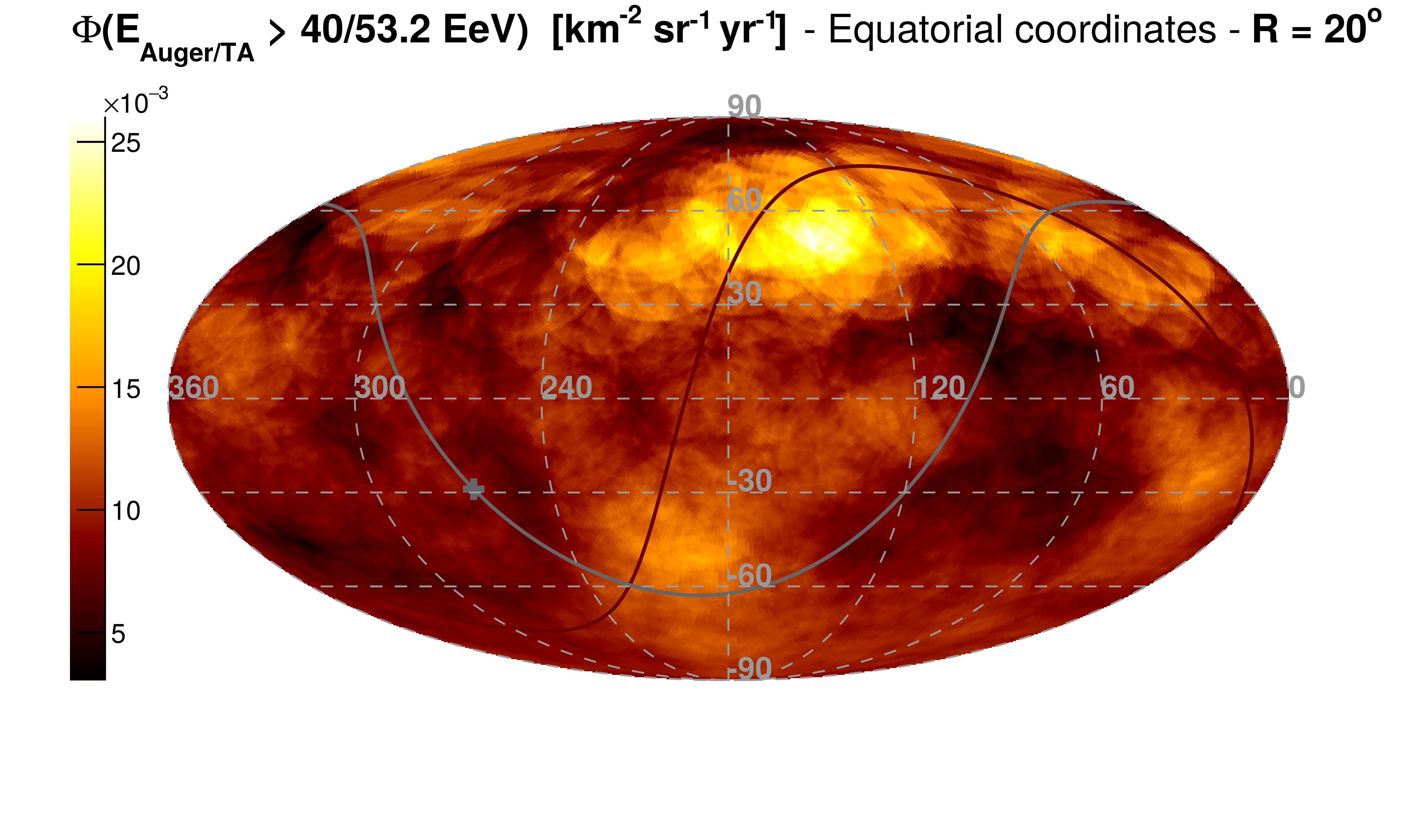}
\includegraphics[width=0.72\textwidth]{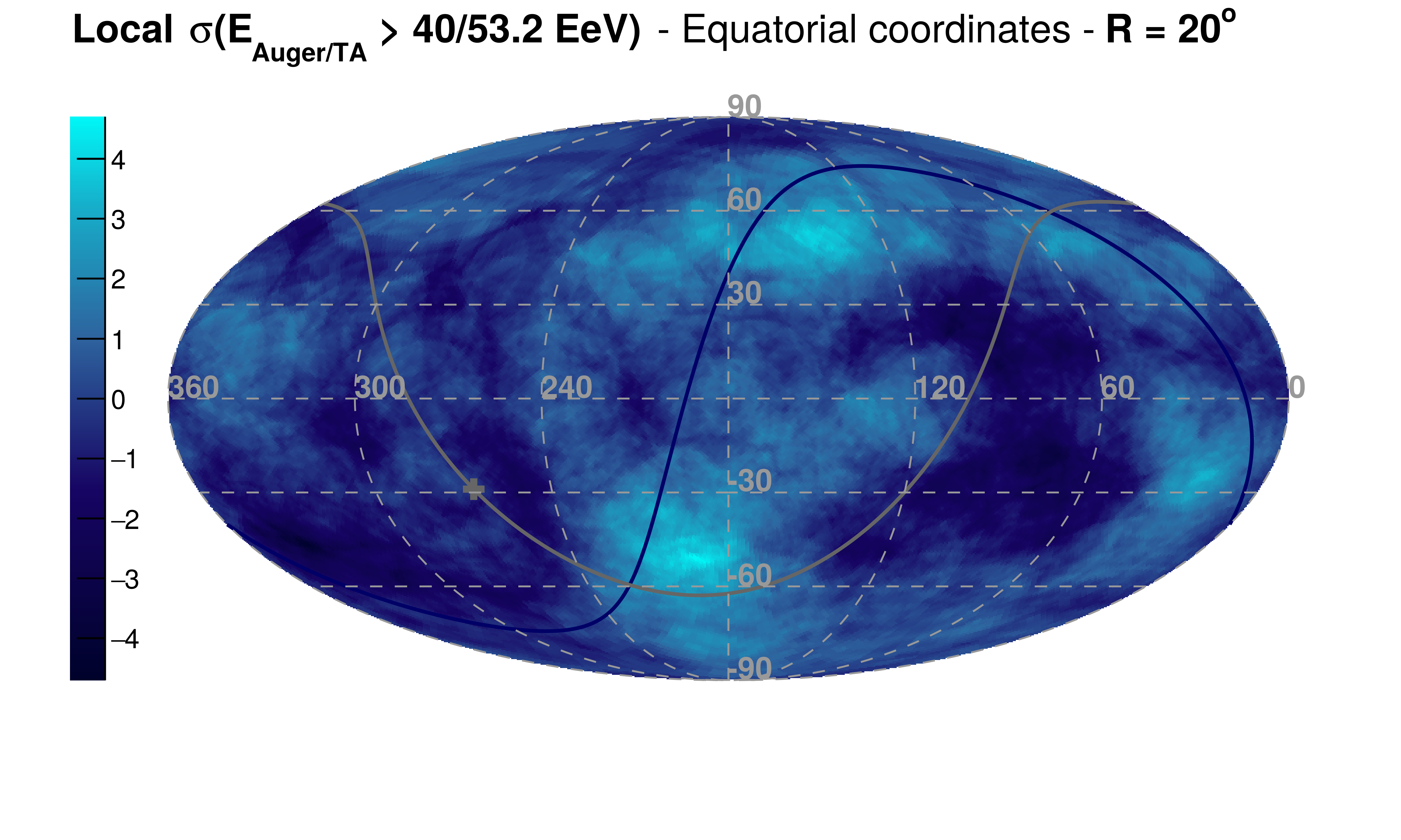}
\includegraphics[width=0.65\textwidth]{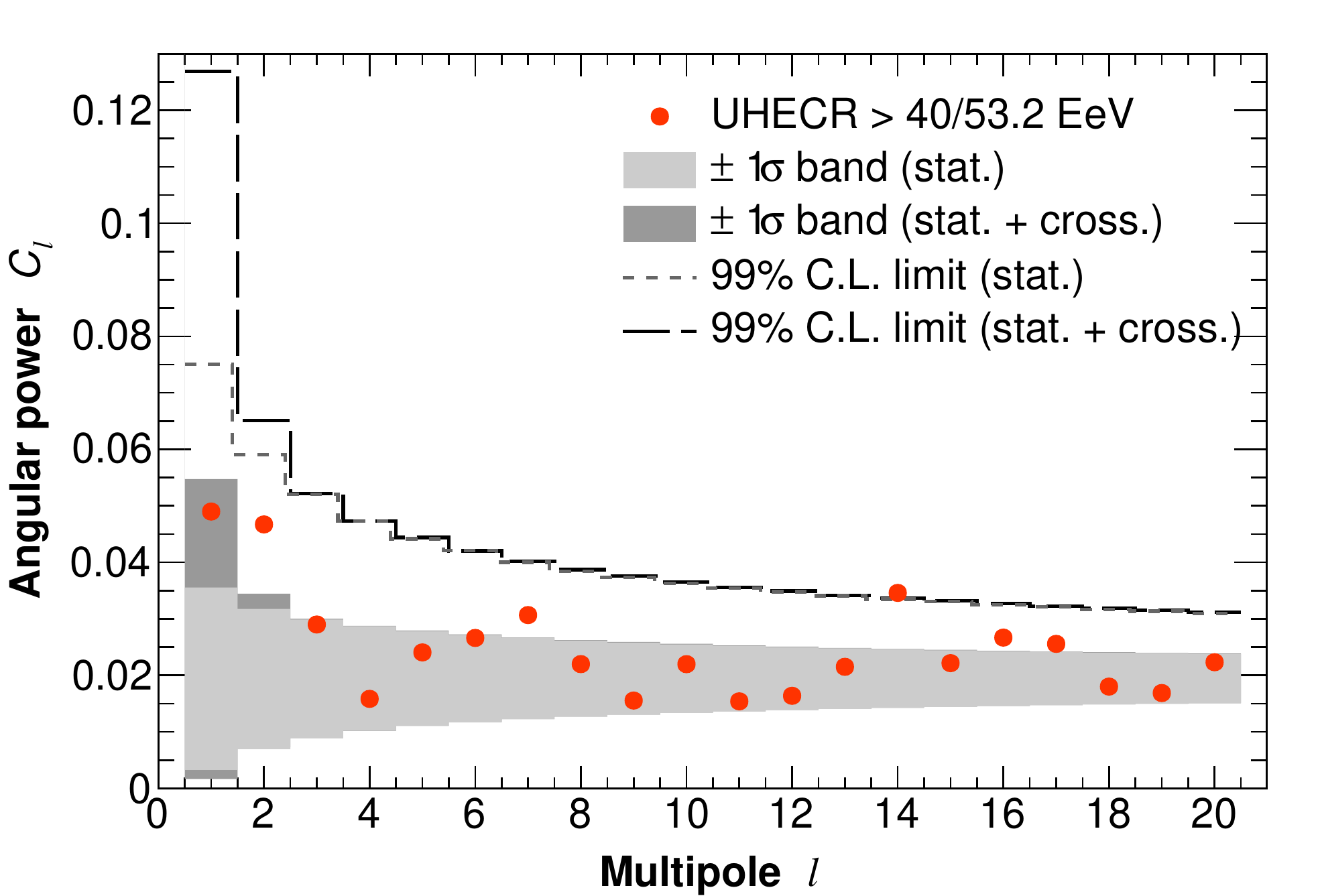}
\caption{Flux ({\it Top}) and significance ({\it Mid-panel}) maps reconstructed at $E_{\rm Auger/TA} > \unit[40/53.2]{EeV}$. The gray line indicates the Galactic plane and the dark line the supergalactic plane. {\it Bottom:} Angular power spectrum at $E_{\rm Auger/TA} > \unit[40/53.2]{EeV}$. See caption in Fig.~\ref{fig-maps-lowE} for more details.}
\label{fig-maps-highE}       
\end{figure*}

The flux and significance maps reconstructed at $E_{\rm Auger/TA} > \unit[40/53.2]{EeV}$ using $20\,\dgr$-radius windows are presented in Fig.~\ref{fig-maps-highE}. Most notably, possible flux enhancements appear along the supergalactic plane, particularly at declinations $\delta \sim \pm\,50\,\dgr$ for the most significant ones.

The angular power spectrum estimated from observations at $E_{\rm Auger/TA} > \unit[40/53.2]{EeV}$ is shown in Fig.~\ref{fig-maps-highE}, bottom. The largest deviation from isotropy (local $\unit[2.8]{\sigma}$) is obtained for a multipole at $\ell=14$, corresponding to an angular scale of about $180\,\dgr/l \sim 13\,\dgr$. Given that no specific angular scale is preferred {\it a priori} at these energies, a penalization for a search over 20 angular scales, or 20 multipoles, results in a post-trial deviation from isotropy at the $\unit[1.6]{\sigma}$ level.

\section{Comparison with the results from each Collaboration}
\label{interpretation}

\subsection{Dipolar component beyond the ankle}

Constraints on the dipolar component reconstructed at  $E_{\rm Auger/TA} > \unit[8.86/10]{EeV}$  are summarized in Table~\ref{tab-dipole}. As also shown in Fig.~\ref{fig-maps-lowE}, bottom, no significant deviation from isotropy is obtained from the 3D analysis presented in this work. Nonetheless, the most likely amplitude of the dipolar component along the Earth rotation axis and along the perpendicular plane, as well as associated uncertainties, are provided in Table~\ref{tab-dipole} for the sake of comparison with previous studies.

Constraints from Auger data only, presented in the first two lines of  Table~\ref{tab-dipole}, were derived in \cite{2017Sci...357.1266P,2018ApJ...868....4A} from a 2D Rayleigh analysis as a function of right ascension and azimuth. Because of partial sky coverage, constraints on the dipole are obtained assuming either no power beyond $\ell=1$, or beyond $\ell=2$. The larger number of effective free parameters in the 3D analysis followed in this work results in a milder deviation from isotropy than that obtained from a 2D study as a function of right ascension. 

A comparison of the central value reconstructed for the component along the Earth rotation axis shows a very good agreement between the 3D and 2D reconstructions. Notably, the limiting factor for the 3D analysis remains the number of events in the cross-calibration band. A somewhat larger tension can be inferred from the estimated central values for the perpendicular component reconstructed in this work and in \cite{2017Sci...357.1266P}. This tension is nearly entirely alleviated when allowing for the presence of power in the quadrupolar component at $\ell=2$ \cite{2018ApJ...868....4A}. Interestingly, the presence of a small-amplitude quadrupolar component could be expected if the sources of UHECRs followed the distribution of local extragalactic matter ({\it e.g.}\ \cite{2018MNRAS.476..715D}).

\subsection{Search for overdensities above the flux suppression}

\begin{figure}
\centering
\includegraphics[width=0.49\textwidth]{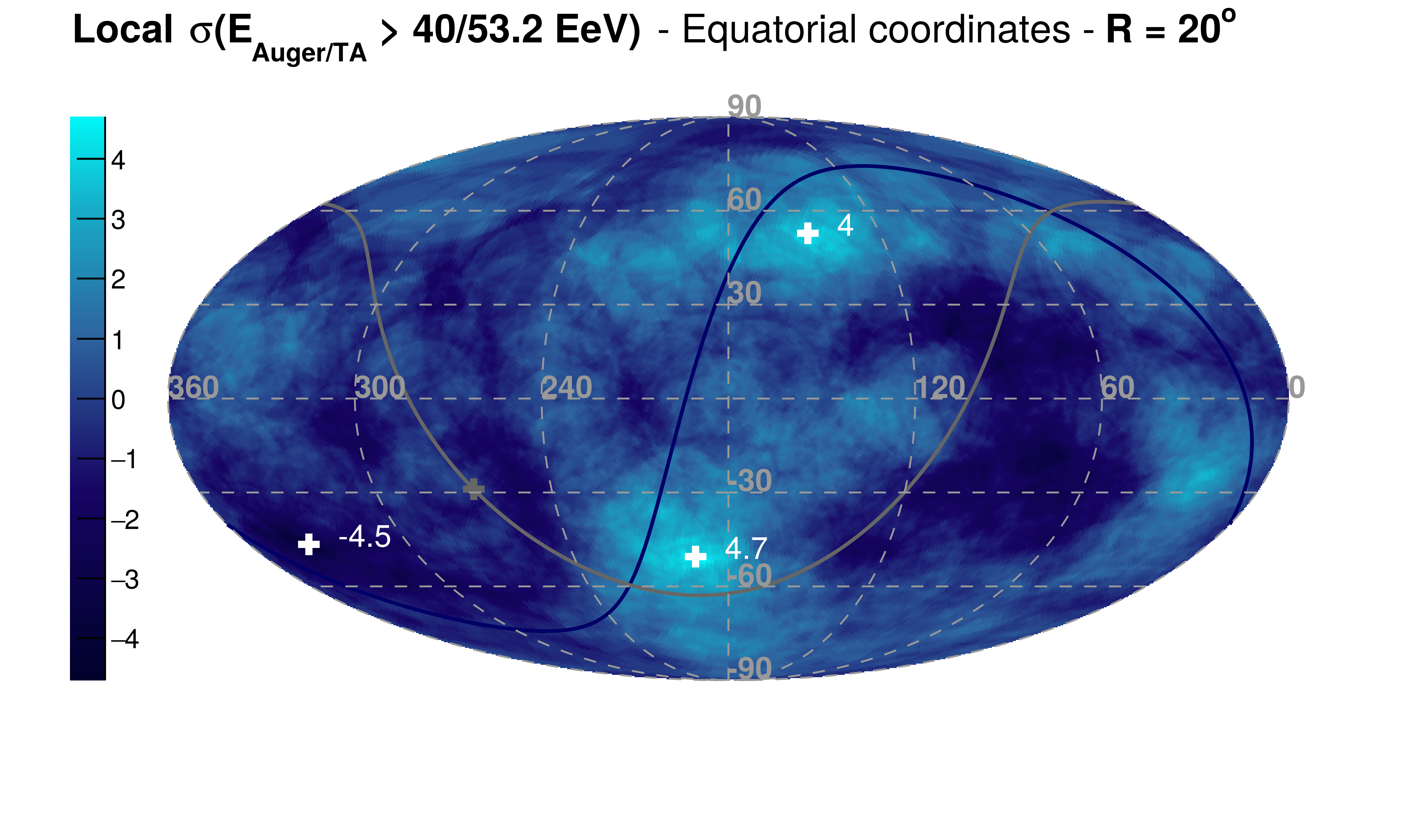}\hfill\includegraphics[width=0.49\textwidth]{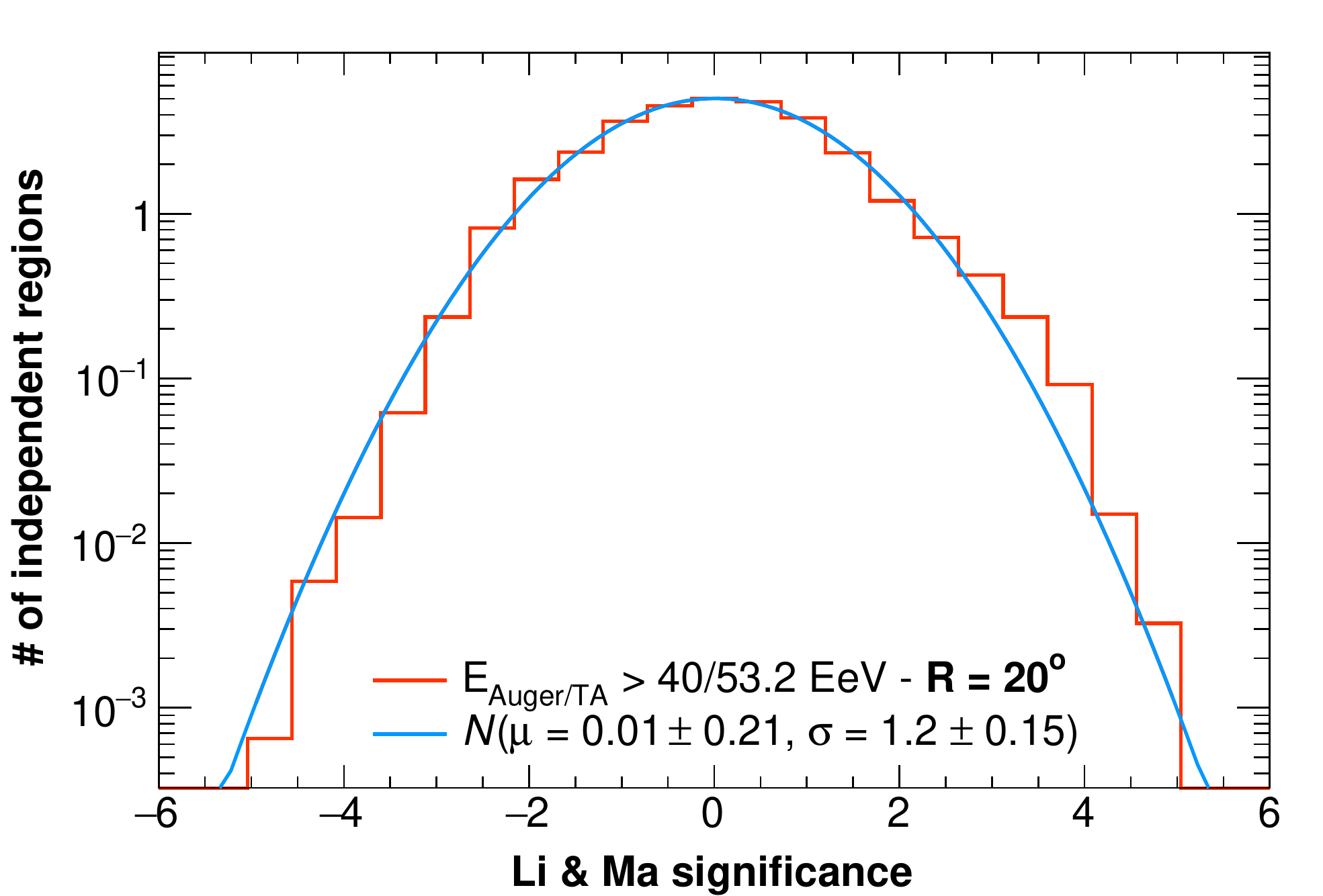}
\caption{{\it Top:} Significance map reconstructed at $E_{\rm Auger/TA} > \unit[40/53.2]{EeV}$ with a $20\,\dgr$ search radius. White crosses indicate deviations from isotropy with ${>}\,4\,\sigma$ local significance. {\it Bottom:} Distribution of the Li \& Ma significance over the sphere.}
\label{fig-overdensities-highE-20deg}       
\end{figure}

\begin{figure}
\centering
\includegraphics[width=0.49\textwidth]{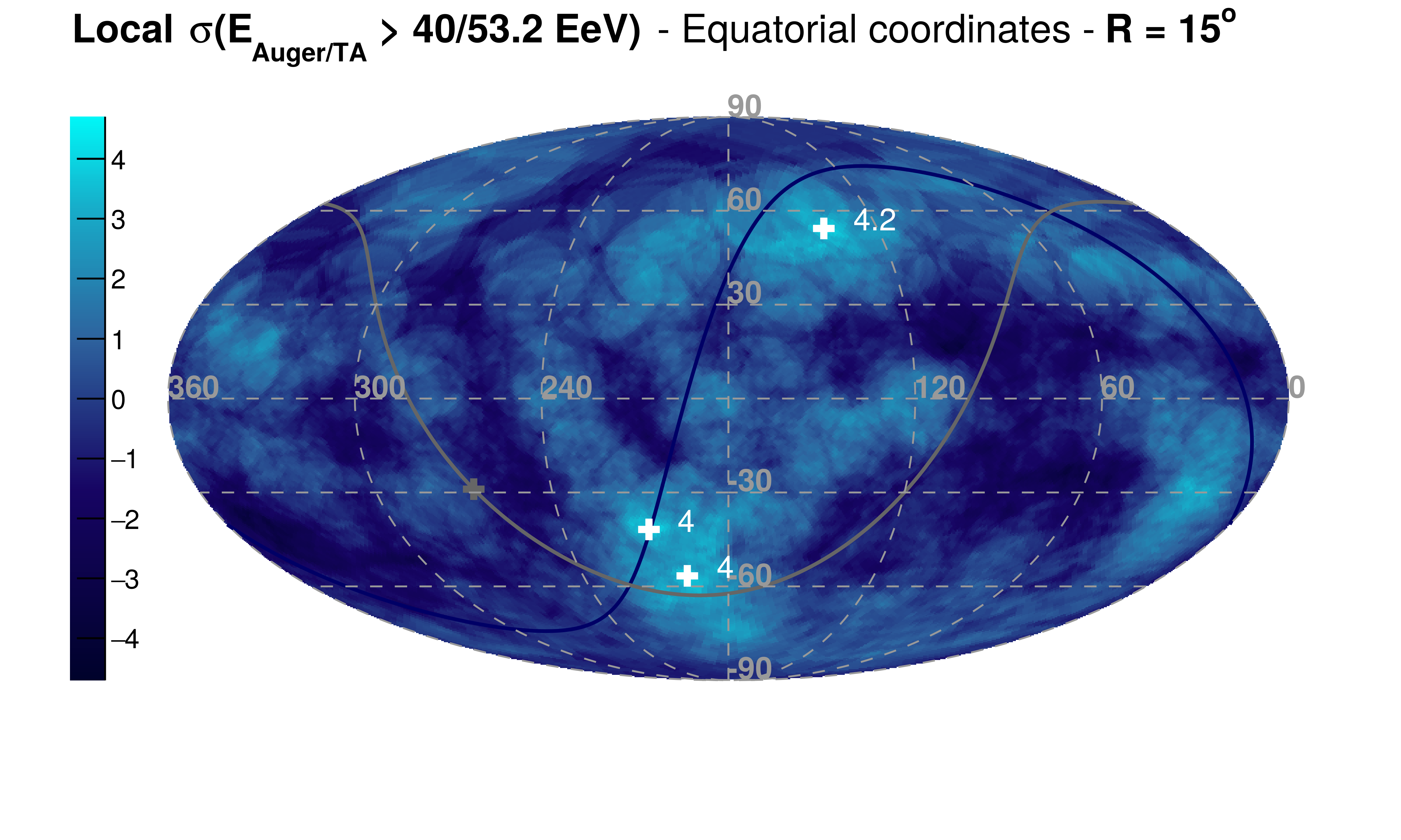}\hfill\includegraphics[width=0.49\textwidth]{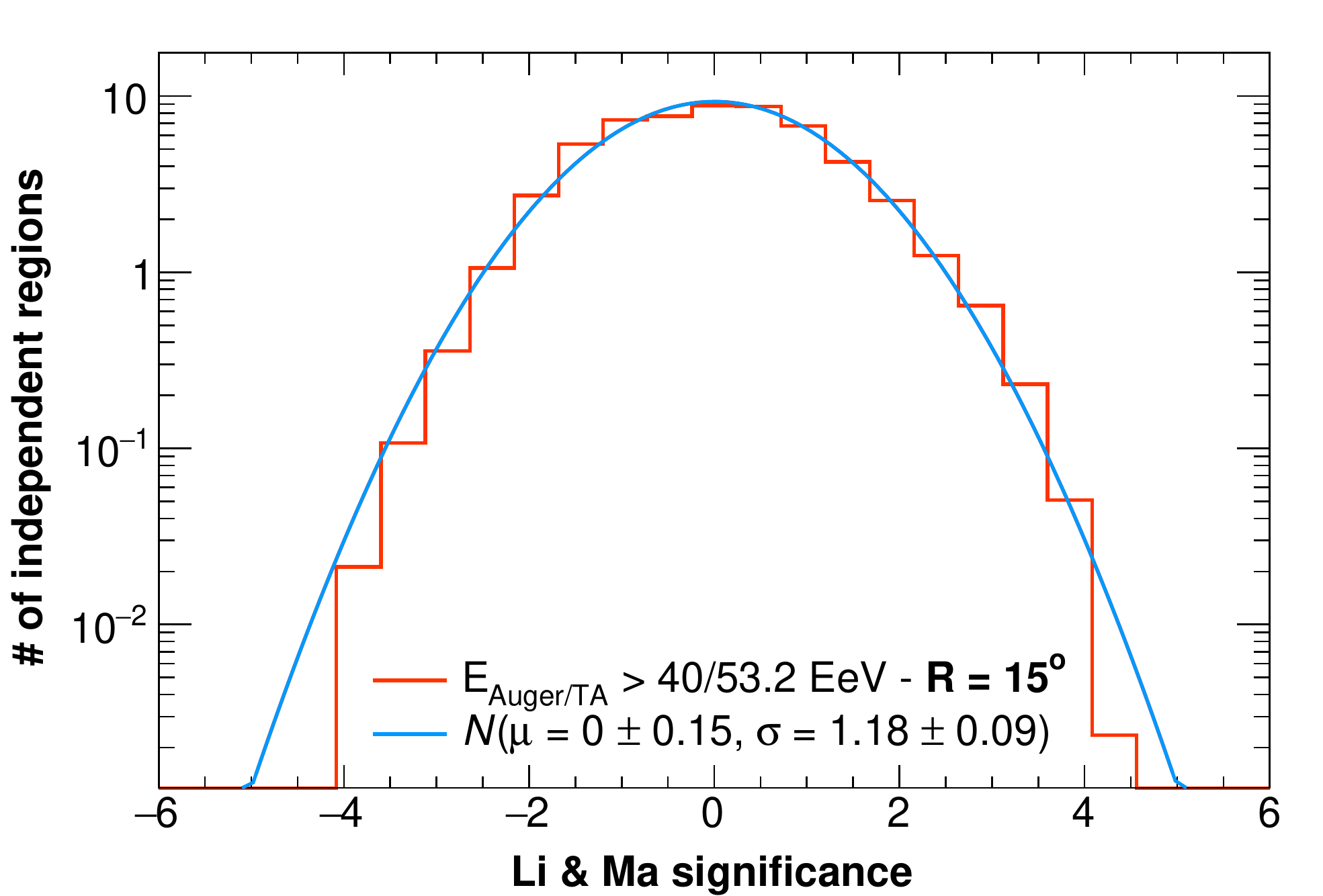}
\caption{{\it Top:} Significance map reconstructed at $E_{\rm Auger/TA} > \unit[40/53.2]{EeV}$ with a $15\,\dgr$ search radius. White crosses indicate deviations from isotropy with ${>}\,4\,\sigma$ local significance. {\it Bottom:} Distribution of the Li \& Ma significance over the sphere.}
\label{fig-overdensities-highE-15deg}       
\end{figure}

Both the Telescope Array and Pierre Auger Collaborations have searched for overdensities in previous studies, with approaches independent of any source model. In \cite{2014ApJ...790L..21A}, the Telescope Array Collaboration performed a search over windows of radius ranging in $15-35\,\dgr$ by steps of $5\,\dgr$, above a fixed energy threshold, $E_{\rm TA} > \unit[57]{EeV}$. The Pierre Auger Collaboration scanned search radii ranging in $1-30\,\dgr$ by steps of $1\,\dgr$, and also scanned energy thresholds $E_{\rm Auger} \in \unit[40-80]{EeV}$ by steps  of $\unit[1]{EeV}$  \cite{2015ApJ...804...15A}. The Northern-sky search resulted in a most-significant excess located at right ascension and declination $(\alpha,\delta) = (146.7\,\dgr, 43.2\,\dgr)$, with a pre-trial significance of $\unit[5.1]{\sigma}$. The largest Southern excess was found at $(\alpha,\delta) = (198\,\dgr, -25\,\dgr)$, with a pre-trial significance of $\unit[4.3]{\sigma}$. In this work, regions of radius ranging in $5-35\,\dgr$ by steps of $5\,\dgr$ are investigated above a fixed energy threshold $E_{\rm Auger/TA} = \unit[40/53.2]{EeV}$. Figures~\ref{fig-overdensities-highE-20deg} and \ref{fig-overdensities-highE-15deg} illustrate the largest deviations from isotropy obtained from this search. 

The most-significant excess is obtained for a $20\,\dgr$ search radius in the direction $(\alpha,\delta) = (193\,\dgr, -50\,\dgr)$, with a local significance of $\unit[4.7]{\sigma}$, while the second one, corresponding to a local significance of $\unit[4.2]{\sigma}$, is obtained with a $15\,\dgr$ search radius in the direction $(\alpha,\delta) = (142\,\dgr, 54\,\dgr)$. These directions are $25\,\dgr$ and $11\,\dgr$ away from the most-significant directions obtained in \cite{2015ApJ...804...15A} and \cite{2014ApJ...790L..21A}, respectively. Given the difference in energy threshold used in this work and those adopted in  \cite{2014ApJ...790L..21A} and \cite{2015ApJ...804...15A}, a difference in peak direction on the order of the preferred search radius is considered as a reasonable agreement.

No straightforward comparison  with previous searches can be made for the local significance of deviation from isotropy: partial- and full-sky-coverage studies were performed above different energy thresholds and thus exploit quite different number of events given the steep decrease of the UHECR spectrum with increasing energy. As shown in Figs.~\ref{fig-overdensities-highE-20deg} and \ref{fig-overdensities-highE-15deg}, the presence of both negative and positive local significances on the order of $\unit[4]{\sigma}$ is indicative of a distribution compatible with isotropy. This can also be inferred from a comparison of the distribution of the significance with a Gaussian distribution: both the mean values and spread estimates are compatible at the $\unit[1-2]{\sigma}$ level with the standard normal distribution expected from isotropy. 

We performed a Monte-Carlo computation of the penalty factor to be applied to the local $p$-values corresponding to the maximum significance reconstructed at $15\,\dgr$ and $20\,\dgr$: it is estimated to be on the order of $10^4$ when accounting for both the scan over the sphere and over the search radius. For comparison, previous search strategies developed independently by each Collaboration would result in penalty factors on the order of $10^3$ and $10^5$ for the energy-independent and energy-dependent searches. The two largest reconstructed local significances, penalized for the procedure adopted in this work, correspond to post-trial significances of $2.2$ and $\unit[1.5]{\sigma}$.

\section{Conclusion}

The Telescope Array and Pierre Auger Collaborations have gathered the largest dataset at ultra-high energies available to date. The combination of data acquired from the Northern and Southern hemispheres enables coverage over the full celestial sphere as demonstrated in previous efforts presented at UHECR conferences. For the first time, studies of both the energy range beyond the ankle and above the flux suppression are presented, where large-scale and intermediate-scale anisotropy patterns could be expected.

The number of events gathered at $E_{\rm Auger/TA} > \unit[8.86/10]{EeV}$ increased by a factor of three with respect to previous effort and by a factor of 1.6 above $\unit[40/53.2]{EeV}$. This large dataset enabled a robust cross calibration of the flux in the common declination band, indicating not only a shift between the energy scales of the two observatories but also a stretch, which has been further investigated by the joint Working Group dedicated to the cosmic-ray spectrum.

While the full-sky model-independent studies presented in this work did not unveil significant anisotropies, either because of the limited number of events in the common declination band above $\unit[8.86/10]{EeV}$ or because of the blind-search approach above $\unit[40/53.2]{EeV}$, interesting leads could be further pursued. This is particularly the case for constraints on the quadrupolar components in the low-energy band and for searches along the supergalactic plane at the highest one, where the presence of a ``ring of fire'' could possibly be inferred from the flux and significance maps shown in Sect.~\ref{maps-highE}.

Future searches for anisotropies based on the dataset presented in these Proceedings may shed new light on the distribution of UHECRs over the celestial sphere, be it on large scales beyond the ankle, where the cumulative flux from sources up to 1\,Gpc could contribute, or on intermediate angular scales at higher energies, where contributions from a few objects following local extragalactic structures could stand out. The hunt for the most extreme accelerators in the Universe along the anisotropy trail continues.

\end{document}